# Title: Unveiling excitonic properties of magnons in a quantum Hall ferromagnet

**Authors:** A. Assouline[1]†, M. Jo[1]†, P. Brasseur[1], K. Watanabe[2],T. Taniguchi[2], T.Jolicoeur[3], P. Roche[1], D.C. Glattli[1], N. Kumada[4], F.D. Parmentier[1] and P. Roulleau[1]*

**Affiliations:**

[1] SPEC, CEA, CNRS, Université Paris-Saclay, CEA Saclay, 91191 Gif sur Yvette Cedex France

[2] National Institute for Materials Science, 1-1 Namiki, Tsukuba, 305-0044, Japan

[3] IPhT, CEA, CNRS, Université Paris-Saclay, CEA Saclay, 91191 Gif sur Yvette Cedex France

[4] NTT Basic Research Laboratories, NTT Corporation, 3-1 Morinosato-Wakamiya, Atsugi 243-0198, Japan

**Magnons enable transferring a magnetic moment or spin over macroscopic distance. In quantum Hall ferromagnet, it has been predicted in the early 90s[1] that spin and charges are entangled, meaning that any change of the spin texture modifies the charge distribution. As a direct consequence of this entanglement, magnons carry an electric dipole moment. Here we report the first evidence of the existence of this electric dipole moment in a graphene quantum Hall ferromagnet [2,3] using a Mach-Zehnder interferometer as a quantum sensor. By propagating towards the interferometer through an insulating bulk, the magnon electric dipole moment modifies the Aharonov-Bohm flux through the interferometer, changing both its phase and its visibility. In particular, we relate the phase shift to the sign of this electric dipole moment, and the exponential loss of visibility to the flux of emitted magnons. Finally, we probe the emission energy threshold of the magnons close to filling factor ν=1. Approaching ν=0, we observe that the emission energy threshold diminishes towards zero, which might be linked to the existence of gapless mode in the canted-antiferromagnetic (CAF) phase at ν=0 [4-6]. The detection and manipulation of magnons based on their electric dipole open the field for a new type of coherent magnon quantum circuits that will be electrostatically controlled.**

Full spin polarization is essential for applications in spintronics, which led, for instance, to the development of a new type of ferromagnetic materials, so-called half-metallic alloys that show complete spin polarization at the Fermi level [7]. In the 90s, it has been demonstrated that a fully polarized system can also be obtained in 2D materials in the quantum Hall effect (QHE) regime. More precisely, in the N=0 Landau level, the kinetic energy is frozen by the magnetic field and all states are degenerate. At the Landau-level filling factor ν=1, the Coulomb energy $E_c = e^2/(\varepsilon l_B)$, much larger than the Zeeman energy $E_Z = g\mu_B B$, plays a primary role and leads to a fully antisymmetric spatial part of the wave function. Therefore, the spin part is symmetric and the ground state is an ideal ferromagnet [8], fully spin-polarized, described by a simple Slater determinant. In the bulk, excitations of this quantum Hall ferromagnet (QHF) are itinerant deformations of the uniform background magnetization known as magnons. These excitations are charge-neutral composed of an electron-hole pair, or exciton, separated by a distance $l_{eh} = k l_B^2$ [9], where k is the magnon wavevector which depends on the magnon energy, and $l_B = \sqrt{\hbar/eB}$ is the magnetic length. This feature relies on a unique property of QHF where the dynamics of spin and charge are entangled, and any modification of the spin texture leads to an electronic density change. Magnon generation is then necessarily tied to an electric dipole moment propagation.

Graphene, which shows a rich sequence of quantum Hall plateaus as spin and valley degeneracies are lifted in the low-lying Landau levels [10], is an ideal platform to study collective excitations at filling factor ν =1[11,12]. Recent experiments [6,13,14] have demonstrated magnon generation by an out-of-equilibrium occupation of edge channels. The magnon detection relies on the absorption at the local vicinity of ohmic contacts. The threshold energy to excite these propagating collective modes is typically $E_Z = g\mu_B B$. However, since the detection relies on the magnon dissipation in an ohmic contact, neither spin nor electrical dipole properties of these collective excitations have been addressed.

In this study, we use the spin-charge entanglement at ν =1 in graphene to probe the excitonic nature of magnons. We first study the reflection properties of magnons at the interface of a PN junction. We then tune such a PN junction into a Mach-Zehnder interferometer (MZI) [15,16,17] and measure the evolution of the phase and visibility of the interferences when magnons are emitted. We show that magnons induce dephasing which cannot be accounted for by the magnetic moment carried by the magnon but only by its electric dipole moment. We also show that loss of visibility can be accounted for by a very simple model, in which decoherence exponentially decreases with the flux of emitted magnons. Furthermore, as the interferometer is defined on the whole width of the sample, it detects all the magnons propagating from one side of the sample to the other. We finally study the threshold energy emission of the magnons close to filling factor $\nu = 1$ and observe hallmarks of the $\nu = 0$ canted-antiferromagnetic (CAF) state.

The sample is in the QHE regime, where the current propagates along edge channels as depicted in Fig. 1A. A top gate enables to tune the filling factor (noted as $\nu_T$) below this latter (filling factor outside the top gated region is noted as $\nu_B$). Due to contact doping, local filling factor close to the ohmic contacts is ν =2 [11]. Applying a voltage $V_E$ on the emitter contact (noted as E in Fig. 1A), a chemical potential difference $\mu = -eV_E$ is set between the inner and the outer edge states (respectively depicted in red and in blue in the figure). When $|eV_E| > E_Z$ ($E_Z \sim 1$ meV at 9 T), spins can flip between the two edges and magnons are emitted: emitter site are noted as e+ for positive $V_E$ and e- for negative $V_E$ (Fig. 1A). Emitted magnons can be detected either by a non-local voltage signal $\frac{dV_i}{dV}$ or phase and visibility change in the MZI where $V$ is the lock-in excitation applied on the emitter contact and $V_i$ the non-local voltage on the $i_{th}$ detector contact.

We first study the non-local voltage properties as a function of $\nu_T$ and $\nu_B$. Following [13], the absorption of a magnon on contact $i = 0,1,2$ creates a chemical potential shift $\varepsilon_i$ in the edge channel flowing from this contact. Similarly, the chemical potential shift generated by the absorption on the grounded contact upstream of contact $i$ creates a chemical potential shift $\varepsilon_{i,G}$ (see Fig. 1A). On contact 0, we thus have $\frac{dV_0}{dV} = (\frac{d\varepsilon_0}{d\mu} - \frac{d\varepsilon_{0,G}}{d\mu})$ where $\mu = -eV_E$ [13]. Figure 1C shows $\frac{dV_0}{dV}$ as a function of the top gate voltage $V_{top}$ ($\nu_T$) at $\nu_B = 1$. At $\nu_T$=1, we detect $\frac{dV_0}{dV}$ above $V_E$=1 mV, a non-local signal between the emitter contact and the contact 0. For $\nu_T$=0, $\frac{dV_0}{dV}$ strongly increases due to the reflected magnons at the PN junction interface that are absorbed in contact 0. Interestingly, this signal remains constant as $\nu_T$ is set from 0 to -1, showing that magnons are still reflected on the $(\nu_T, \nu_B) = (-1,1)$ interface. A straightforward explanation for this strong reflection would be that the magnon velocity mismatch between $\nu_B$=1 and $\nu_T$=0 prevents magnons from being transmitted across the $(\nu_T, \nu_B) = (0,1)$ interface. In that case, one can assume that a thin strip at $\nu_T = 0$ develops along the interface at $(\nu_T, \nu_B) = (-1,1)$, thereby also reflecting magnons in this configuration.

Measuring simultaneously $\frac{dV_1}{dV}$ and $\frac{dV_2}{dV}$, we observe a remarkable property of these magnons. At the edge of $\nu_T=1$, the non-local voltage decrease of $\frac{dV_1}{dV}$ corresponds to a sharp increase of $\frac{dV_2}{dV}$ what means that magnons are deflected from contact 1 to contact 2. We argue that the deflection in the bulk is due to the magnons excitonic nature. Indeed, magnons in the QHF carry an electric-dipole moment $\mathbf{p}=|e|l_B^2 \hat{\mathbf{z}} \times \mathbf{k}$, with $\mathbf{k}$ the center of mass momentum and $\hat{\mathbf{z}}$ collinear to the magnetic field [9] (yellow arrow in Fig. 1A). This electric-dipole interacts with any charged localized states that typically form in the bulk of the sample.

To explore the excitonic nature of magnons, we now tune the sample in a bipolar quantum Hall state to form a MZI as described in our previous study [18]. In the N region, the filling factor is $\nu_N = 1$ and one spin-up channel circulates anticlockwise along the boundary, while in the P region $\nu_P = -2$ and two channels of the opposite spin (↑,↓) circulate clockwise (Fig. 2A). By applying side gates at the intersections between the physical edge of graphene and the PN interface, we can realize two valley splitters that insure coherent mixing of the copropagating but opposite-valley-isospin channels of the PN interface. In the $\nu_N = 1$ and $\nu_P = -2$ configuration, a MZI is formed at the PN junction interface defined by the two spin-up channels. A DC bias $V_E$ is applied on the emitter contact E and the electronic transmission probability $T_{MZ} = I_T / I_0$ (with $I_0$ the injected current and $I_T$ the transmitted current through the PN junction measured in contact T) is measured as a function of the magnetic field and $V_E$ (in Fig. 2B). At the threshold energy $|eV_E| \sim E_Z$, we observe a clear phase shift of $T_{MZ}$ that is increasing with the applied DC bias, accompanied by a drop in the oscillations visibility V, as shown in Fig. 2C. A slight change of the magnetic field (from 8.875T to 8.85T) can dramatically alter the phase shift, but not the visibility drop. As seen in Fig 2D and 2E, the phase shift is now very asymmetric and non-monotonous. This definitively rules out the phase shift induced by the magnetic moment carried by the emitted magnons which does not depend on the sign of $V_E$ and the magnetic field.

Instead, we argue that both the phase shift and visibility drop are due to the excitonic nature of the magnons. Indeed, as depicted in Fig. 2A, depending on their incidence angle with respect to the PN junction, the dipole carried by the magnons acts as a local electrostatic gate, affecting the AB phase of the MZI. This phase is particularly sensitive to the projection $p_\perp$ of the magnon electric dipole perpendicular to the junction (along the direction labeled $x$ in Fig. 2A), increasing (resp. decreasing) when the projection is positive (resp. negative). Assuming that $n$ magnons impinge on the junction, we write $p_{i,\perp}^+$ (resp. $p_{i,\perp}^-$) when the electrical dipole moment projection along x is positive (resp. negative). The sum of all the positive (resp. negative) contribution is noted $I_{M,+} = \sum_{i=0}^{m} p_{i,\perp}^+$ (resp. $I_{M,-} = \sum_{j=0}^{p} p_{i,\perp}^-$) with $p + m = n$. The phase shift induced by magnons noted as $\varphi_M$ will then depend on the total electric dipole: $\varphi_M \propto I_{M,+} - I_{M,-}$. This additional phase term will modify the final state into $|\Psi_{final}\rangle = r_1|\downarrow,\vec{w}\rangle + t_1 e^{i(\varphi_{AB}+\varphi_M)}|\downarrow,-\vec{w}\rangle$ with $r_1$ and $t_1$ the reflection and transmission probability of the first valley splitter, $\varphi_{AB}$ the AB phase and $\vec{w}(-\vec{w})$ the valley states of the two copropagating edge states along the PN junction interface [18]. The transmission probability through the MZI reads $T_{MZ} = |r_1 t_2|^2 + |r_2 t_1|^2 + 2|r_1 r_2 t_1 t_2|\cos(\varphi_{AB} + \varphi_M)$, directly reflecting the phase shift induced by magnons (Figs. 2B and 2D). The strong visibility drop can in turn be understood by the fact that the phase $\varphi_M$ fluctuates as magnons randomly impinge on the junction. Assuming a Poissonian emission of magnons, the fluctuations of $\varphi_M$, depends on the total amount of propagating magnons: $\langle \delta\varphi_M^2 \rangle \propto |I_{M,+}| + |I_{M,-}|$, with $|I_{M,+}|$ and $|I_{M,-}|$ proportional to $|V_E| - E_z/e$. The oscillating term in the above transmission $T_{MZ}$ then acquires an additional prefactor $e^{-\frac{\langle \delta\varphi_M^2 \rangle}{2}}$, yielding an exponential decrease in the MZ visibility as a function of $|V_E| - E_z/e$.

In Fig. 2F, we plot ln(V) with V the visibility shown in Fig. 2E. The perfect agreement with a linear dependence (red solid line in Fig. 2F) validates our assumption of a Poissonian source of magnons. The phase shift being close to 0, this means that $I_{M,+} \sim I_{M,-}$. At lower magnetic field (in Fig. 2C) the phase dependence with $V_E$ is much more pronounced. We compare the dependence of both ln(V) and the phase shift as a function of $V_E$ in Fig. 2G, showing a perfect match between the two at our experimental accuracy. This can be explained if magnons predominantly impinge on the junction from a given direction. The sign of the phase shift indicates that the area of the interferometer is getting smaller when magnons are emitted and the projection of the electrical dipole moment of the emitted magnons along $x$ is mainly positive. Assuming $I_{M,+} \gg I_{M,-}$, this leads to $\langle \delta\varphi_M^2 \rangle \propto |I_{M,+}|$ and $\varphi_M \propto I_{M,+}$. Decoherence and phase shift will thus have the same dependence on $V_E$, in agreement with our measurement in Fig. 2G. The comparison between Fig. 2C and Fig. 2E shows that, on the one hand, the visibility only weakly depends on the magnetic field, implying that the total flux of magnons impinging on the junction remains roughly constant. On the other hand, the phase shift changes dramatically when modifying the field of 25mT, indicating that the magnon spatial distribution is very sensitive to the magnetic field.

It could be argued that the observed phase shift and visibility drop in the MZI oscillations are caused by the absorption of magnons on a contact upstream of the interferometer. The resulting dc bias $V_{inj}$ developing across the MZI could indeed be a source of phase shift and decoherence [18,19,20]. This dc bias is obtained by measuring the non-local differential voltage across the MZI $\frac{dV_T}{dV}$, where $V_T$ is the non-local voltage developing on the contact on which $T_{MZ}$ is measured, and $V$ the voltage applied on the magnon emission contact (see e.g. Fig. 2A). For a finite $V_E$ and a given transmission $T_{MZ}$ (measured independently), we have $V_{inj} = \frac{1}{T_{MZ}*e} \int_0^{eV_E} \frac{dV_T}{dV}(\varepsilon) d\varepsilon$. At $V_E$=1.5 mV, we compute $V_{inj}$ =1 µV. We have recently shown that the visibility of MZI obtained in graphene PN junctions remains unaffected by the dc bias up to values much larger than the computed $V_{inj}$ [18]. We show in Fig. 3 that it is also the case here, with MZI oscillations, plotted as a function of the magnetic field and dc bias $V_I$ (see Fig. 3A), surviving up to $V_I \sim 250$ µV (Fig. 3B). This is further confirmed by computing the visibility, which shows the lobe structure typical of MZ interferometers over scales of 100 µV (Fig. 3C), as well as the phase, which also changes over voltage scales 2 orders of magnitude larger than the computed $V_{inj}$ (Fig. 3D). The dephasing and decoherence effects observed when magnons are emitted thus only result from the electric-dipole moment of the magnons interacting with electrons in the MZI.

In the last section, we discuss how the magnons emission energy changes with the filling factor around $\nu_B = 1$ plateau. Fig. 4A shows the Hall resistance below the back gate around $\nu_B = 1$ as a function of the back-gate voltage $V_{BG}$. The MZ oscillations are measured for different $V_{BG}$, shown as vertical dashed lines. On the edges of the plateau (blue and black vertical dotted lines), the oscillations differ markedly. Indeed, close to $\nu_B = 2$ (Fig. 4B, blue vertical line in Fig. 4A), the visibility starts dropping at a value of $V_E$ only slightly larger than in the middle of the plateau (see *e.g* Fig. 2, as well as Fig. 4D) and the phase is weakly affected; close to $\nu_B = 0$ (Fig. 4C, black vertical line in Fig. 4A), on the other hand, the visibility drops drastically, at a much smaller $V_E < 0.5$ mV. The apparent decrease of the magnon emission threshold as the electron density is tuned close to $\nu_B = 0$ was not observed in previous experiments, where the electron density at the emission point was kept constant [13].

At $\nu_B = 0$ the quantum Hall state of graphene also supports four zero-energy Goldstone modes in the approximation of complete SU(4) symmetry [21]. With realistic anisotropies taken into account, the situation depends on the nature of the ground state. If it is the CAF state which is the ground state, then there is exactly one gapless mode all the other ones being gapped [20]. The dispersion relation of this gapless mode is given by: $E_{CAF}(k) = 2k\sqrt{\rho_n(u_\theta + \rho_\theta k^2)}$, where $\rho_{n,\theta}$ are stiffnesses [22] that depend upon the canting angle $\theta$ governed by the ratio of Zeeman energy and anisotropy $\cos(\theta) = \frac{\pi l^2 \mu_B B}{|g_\perp|}$, and $u_\theta$ is the anisotropy energy. Therefore, we expect a smaller threshold magnon emission energy when approaching $\nu_B = 0$.

To conclude, we have used an electronic MZI to study the microscopic structure of magnons in graphene QHF. In particular, we have brought, for the first time, the evidence for the existence of the electric dipole moment associated with magnons. Once emitted by an ohmic contact, magnons propagate towards the interferometer and interact with the electrical field at the interface of the PN junction. Depending in the electrical dipole of magnons AB flux through the interferometer changes, thus inducing the phase shift of the interference. A visibility drop is also observed with the magnon emission demonstrating that they are a source of decoherence. Finally, the emission energy of the magnons are investigated as a function of the filling factor. Zero gapped magnon modes are detected when approaching the $\nu = 0$ CAF mode. This study opens the way to a new type of hybrid circuits where we could envision electrostatically controlled magnons coupled to flying qubits in a coherent manner.

**Acknowledgments:** We warmly thank Heung-Sun Sim, June-Young Lee, Mark O. Goerbig, Chun-Li Huang, Nemin Wei and Allan Mac Donald for enlightening discussions, as well as P. Jacques for technical support.

**Funding:** This work was funded by the ERC starting grant COHEGRAPH, the CEA, the French Renatech program, ``Investissements d'Avenir'' LabEx PALM (ANR-10-LABX-0039-PALM) (Project ZerHall), and by the EMPIR project SEQUOIA 17FUN04 co-financed by the participating states and the EU's Horizon 2020 program. It is also supported by Korea NRF via the SRC Center for Quantum Coherence in Condensed Matter (GrantNo.2016R1A5A1008184).

**Author contributions:**

A.A., MJ., P.B. & P.Rou performed the experiment with help from F.D.P. and P.R.; A.A, P.B., M.J., F.D.P & P.Rou analysed and discussed the data with help from P.R.; Th. J. & P.Rou developed the theoretical model; T.T., K.W. provided the BN layers; M.J. fabricated the device with inputs from W.D., P.B., A.A., F.D.P & P.Rou; A.A and P.Rou wrote the manuscript with inputs from all coauthors; P.Rou designed and supervised the project.

\* corresponding author

† equal contribution

**competing interests:** none declared

**Data and materials availability:** All data, code, and materials used in the analysis are available in some form to any researcher for purposes of reproducing or extending the analysis.

**Methods:**

**Measurements:**

We used a Cryoconcept dry dilution refrigerator with a base temperature of 13mK. Measurements of transmitted currents and $R_H$ values were performed using multiple lock-in amplifiers with low noise preamplifiers. AC excitations 1nA-10nA with different frequencies (70Hz-300Hz) were used. Buried ohmic contacts underneath top gates enabled us the direct determination of filling factors from regions of interest.

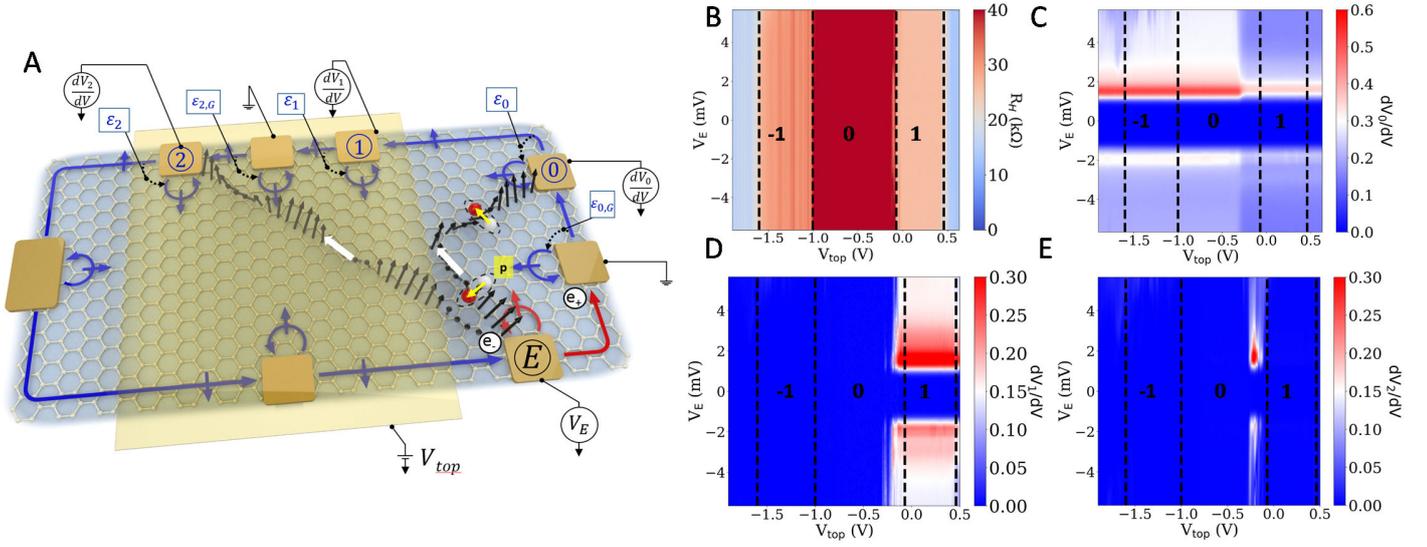

**Fig. 1: Transmission and reflection of magnons at a PN junction interface. (A)** Schematic representation of the device in the quantum Hall regime. A metallic top gate enables to change the filling factor noted as $\nu_T$ below this latter. Filling factor of the region which is not covered by the top gate is noted as $\nu_B$. Local doping near the metallic contacts increases locally the filling factor to ν=2 and this is represented by additional small loops of edge channel on each contact. When the applied DC bias noted as $V_E$ reaches the Zeeman energy $|eV_E|\sim E_Z$, magnons are emitted, which is depicted as groups of black arrows. For positive (negative) voltage the collective excitation emission site is represented by the encircled e₊ (e₋) sign. $\varepsilon_i$ is the redistributed chemical potential at the $i_{th}$ absorption site. **(B)** Measured Hall resistance below the top gate as a function of the top gate voltage $V_{top}$. **(C)** Non-local voltage signal measured in contact 0, $\frac{dV_0}{dV}$, as a function of $V_{top}$ and $V_E$ **(D)**, **(E)** Non-local voltage signal measured in contact 1, $\frac{dV_1}{dV}$, and contact 2, $\frac{dV_2}{dV}$, respectively, as a function of $V_{top}$ and $V_E$.

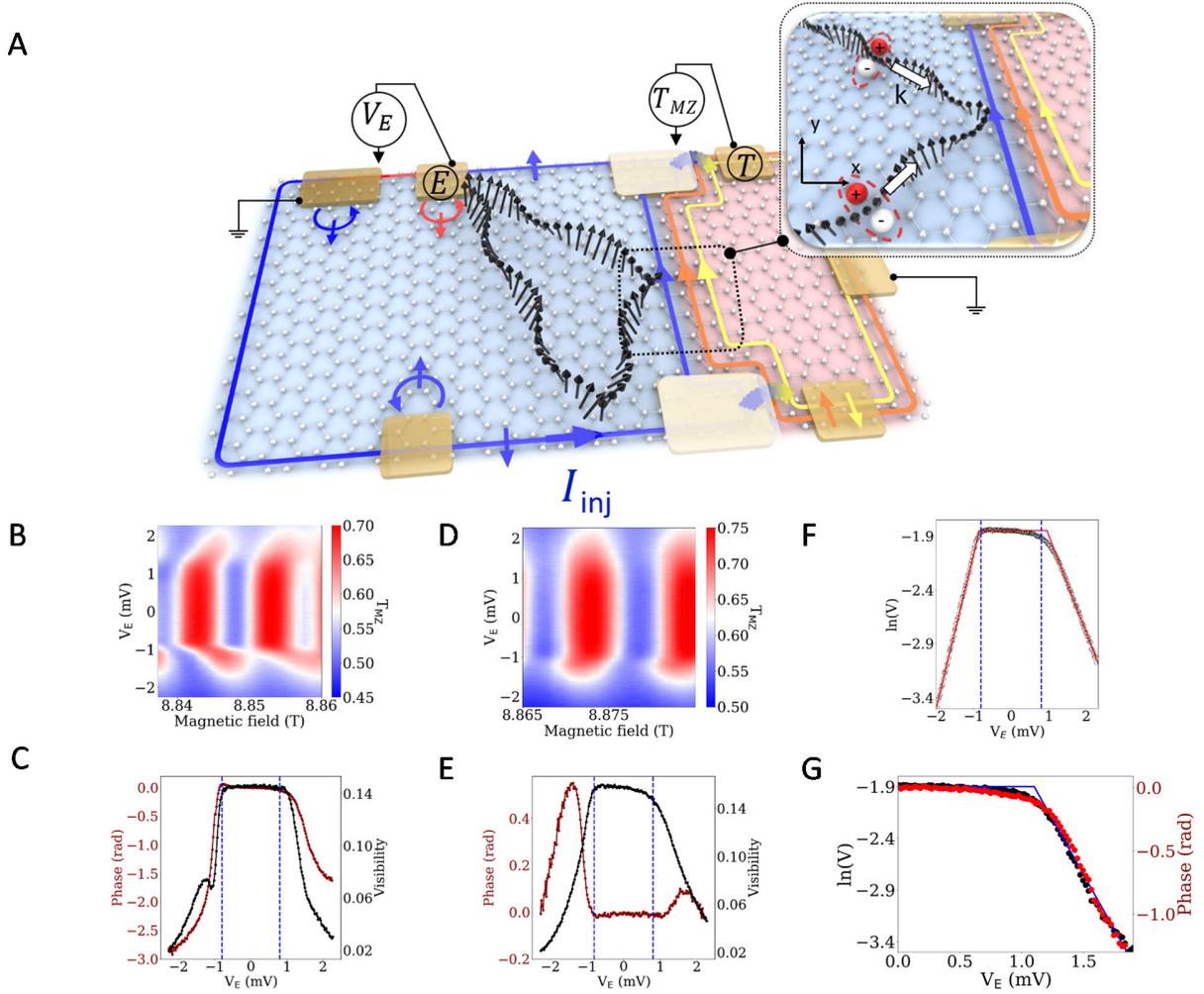

**Fig. 2: Magnons detection with a MZI.** (**A**) MZI defined at the interface of a PN junction. The N region is depicted in blue, the P region is in pink. Electrons are injected from the lower left ohmic contact and the MZI transmission probability, $T_{MZ}$, is measured at the upper right contact. The two spin up edge states define the MZI. Applying $V_E$, magnons are emitted towards the interferometer. **inset**: Schematic representation of the particle-hole pairs associated with the magnons for two different incidence angles. The electric dipole moment is given by $\mathbf{p}=|e|l_B^2 \hat{\mathbf{z}} \times \mathbf{k}$. Depending on the incident angle of the magnons, the electric dipole moment is positive or negative along the x direction. (**B**) $T_{MZ}$ as a function of $V_E$ and the magnetic field between 8.84T and 8.86T. (**C**) Computed visibility and phase from (**B**). (**D**) $T_{MZ}$ as a function of $V_E$ and the magnetic field between 8.865T and 8.885T. (**E**)

Computed visibility and phase from (**D**). **(F)** ln(V) as a function of $V_E$ with V the visibility in 2E. **(G)** ln(V) (black dots) and phase (red dots) as a function of $V_E$ with V and the phase extracted from 2C.

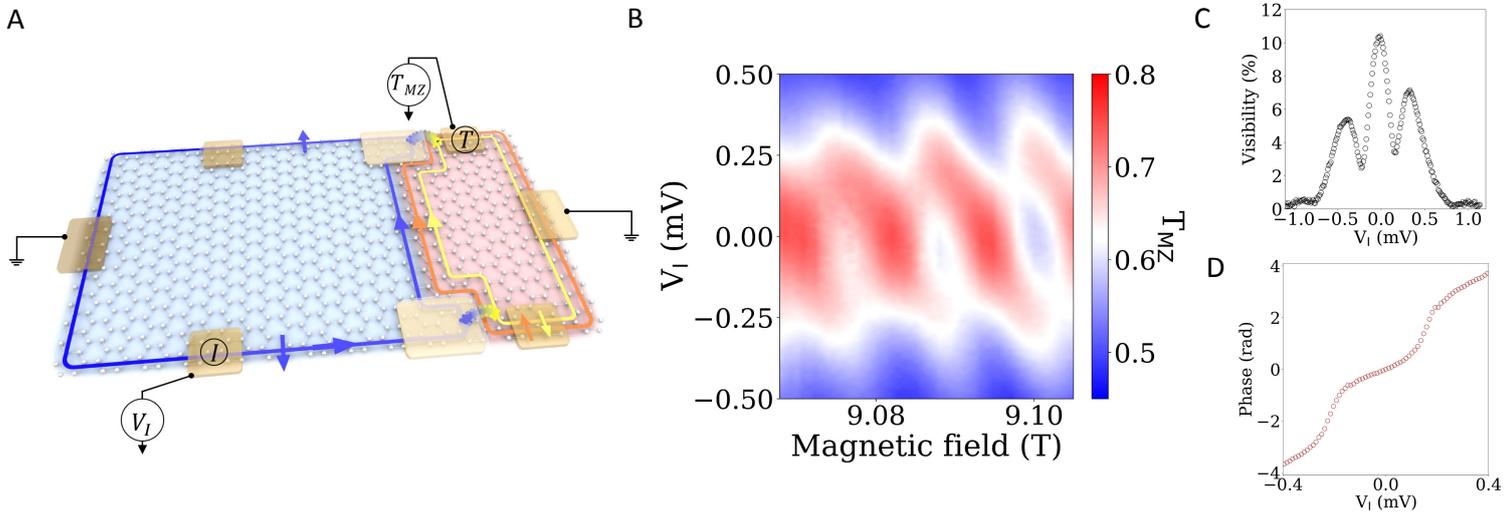

**Fig. 3. Phase shift and decoherence as a function of the injected current into a MZI.** (**A**) Bias noted as $V_I$ is now directly applied on contact I. (**B**) $T_{MZ}$ as a function of the magnetic field and the bias $V_I$. (**C**), (**D**) Visibility and phase of the current oscillations computed from (**B**) as a function of $V_I$, respectively.

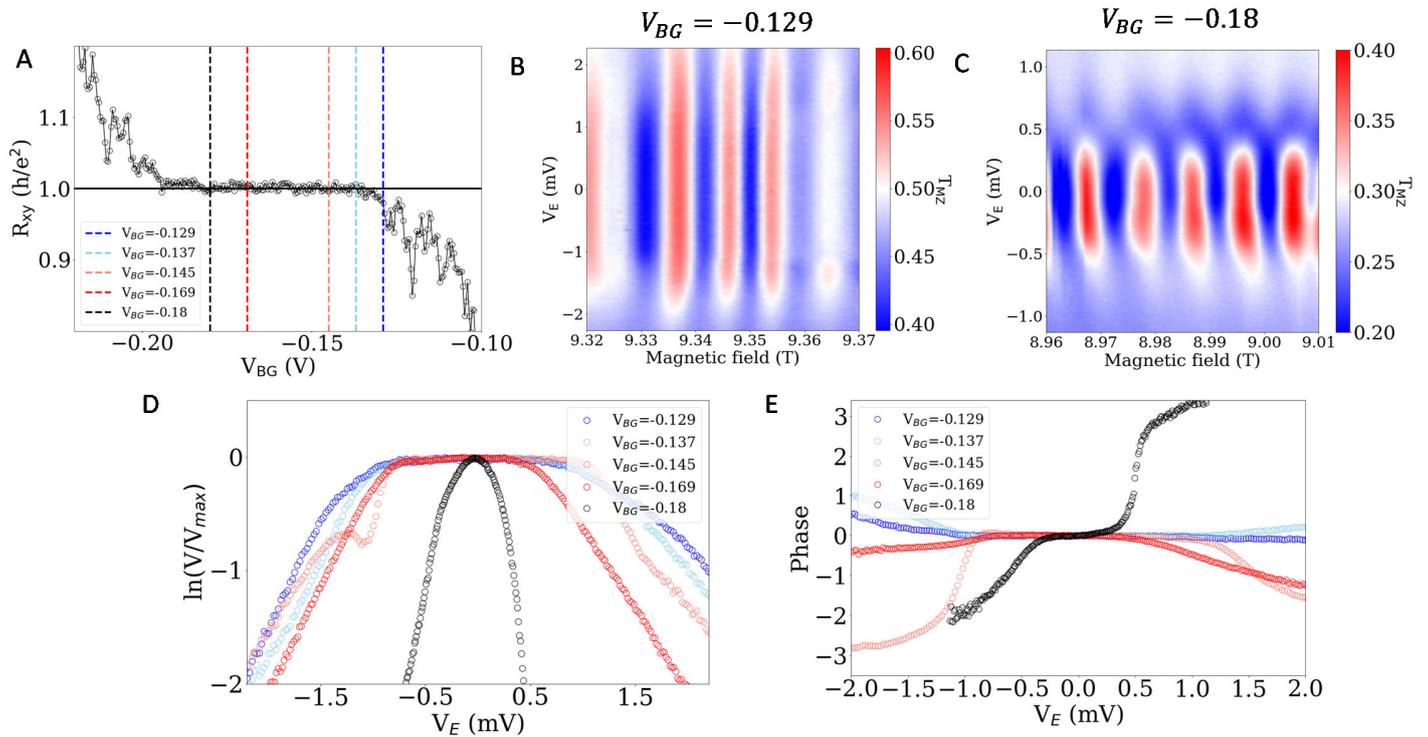

**Fig. 4: Magnon properties around $\nu_B$=1 and when approaching to $\nu_B$=0 or $\nu_B$=2.** **(A)** Two points $R_{xy}$ measurement as a function of the back gate voltage $V_{BG}$. For $V_{BG}$=-0.18V we are on the edge of the $\nu_B$ =1 plateau closer to $\nu_B$ =0 and for -0.129V, the other edge closer to $\nu_B$ =2. **(B), (C)** $T_{MZ}$ as a function of $V_E$ and the magnetic field for $V_{BG} = -0.129V$ and $V_{BG} = -0.18$V, respectively. **(D)** Natural logarithm of the normalized visibility (visibility divided by its maximum) as a function of $V_E$ for different $V_{BG}$. **(E)** Phase of the current oscillations as a function of $V_E$ for different $V_{BG}$.